\documentstyle[12pt,epsf]{article}
\begin{document}

\newcommand{\tstrut}{\vline height11pt depth7pt width0pt}
\def\op{{\cal O}}
\def\lsim{\mathrel{\lower4pt\hbox{$\sim$}}\hskip-12pt\raise1.6pt\hbox{$<$}\;
}
\def\Dd{\psi}
\def\pp{\lambda}
\def\ket{\rangle}
\def\BAR{\bar}
\def\xba{\bar}
\def\fa{{\cal A}}
\def\fm{{\cal M}}
\def\fl{{\cal L}}
\def\ufs{\Upsilon(5S)}
\def\gsim{\mathrel{\lower4pt\hbox{$\sim$}}
\hskip-10pt\raise1.6pt\hbox{$>$}\;}
\def\ufour{\Upsilon(4S)}
\def\xcp{X_{CP}}
\def\ynotcp{Y}
\vspace*{-.5in}
\def\etap{\eta^\prime}
\def\bfb{{\bf B}}

\def\uglu{\hskip 0pt plus 1fil
minus 1fil} \def\uglux{\hskip 0pt plus .75fil minus .75fil}

\def\slashed#1{\setbox200=\hbox{$ #1 $}
    \hbox{\box200 \hskip -\wd200 \hbox to \wd200 {\uglu $/$ \uglux}}}
\def\slpar{\slashed\partial}
\def\sla{\slashed a}
\def\slb{\slashed b}
\def\slc{\slashed c}
\def\sld{\slashed d}
\def\sle{\slashed e}
\def\slf{\slashed f}
\def\slg{\slashed g}
\def\slh{\slashed h}
\def\sli{\slashed i}
\def\slj{\slashed j}
\def\slk{\slashed k}
\def\sll{\slashed l}
\def\slm{\slashed m}
\def\sln{\slashed n}
\def\slo{\slashed o}
\def\slp{\slashed p}
\def\slq{\slashed q}
\def\slr{\slashed r}
\def\sls{\slashed s}
\def\slt{\slashed t}
\def\slu{\slashed u}
\def\slv{\slashed v}
\def\slw{\slashed w}
\def\slx{\slashed x}
\def\sly{\slashed y}
\def\slz{\slashed z}

\rightline{BNL-HET-01/17}
\rightline{Fermilab-Pub-01/104-T}

\begin{center}

%
%
%
%

{\large\bf
Determining the Phases $\alpha$ 
and
$\gamma$
from Direct CP Violation in 
$B_u$, $B_d$ and $B_s$ Decays to Two 
Vectors
}
\vspace{.2in}

David Atwood$^{1}$\\
\noindent Dept. of Physics and Astronomy, Iowa State University, Ames,
IA
50011\\
\medskip

Amarjit Soni$^{2}$\\
\noindent Theory Group, Brookhaven National Laboratory, Upton, NY
11973\\
\footnotetext[1]{email: atwood@iastate.edu}
\footnotetext[2]{email: soni@bnl.gov}
\end{center}
\vspace{.15in}

\begin{quote}

{\bf Abstract:} A method for clean determination of the unitarity angles
$\alpha$ and $\gamma$ is proposed that uses only direct CP violation and
does not require any time dependent measurements.  The method takes
advantage of helicity amplitudes for $B_u$, $B_d$ and $B_s$ decay to two
vector mesons and can be used, at any B-facility, in conjunction with a
large number of modes. It also allows for experimental tests of
theoretical approximations involved.

\end{quote}

Considerable progress has recently been made in experimental determination
of the angle $\beta$ of the unitarity triangle and improved measurements
are expected in the near future~\cite{cdf_ref,belle_ref,babar_ref}.
However, even with increased precision in the extraction of $\beta$, the
ability to test the CKM~\cite{cabibbo} description of CP violation in the
Standard Model (SM) will be limited as the existing tests rely on
theoretical calculation of hadronic matrix elements which still have
considerable uncertainties~\cite{CKMparadgim}. Therefore clean
determinations of all three angles ($\alpha$, $\beta$, $\gamma$) of the
unitarity triangle is important to facilitate precision tests of the SM
and search for new CP-odd phase(s) due to physics beyond the SM.  
Currently the asymmetric B-factories are performing remarkably well and in
the future hadronic B-machines may produce 1-3 orders of magnitude more
B-mesons. Thus, it is very important to devise methods for clean
extraction of the angles that can be used at all B-facilities.
Furthermore, since time dependent measurements involving $B_s$ mesons are
extremely challenging due to the high frequency of $B_s-\bar B_s$
oscillations, it would also be helpful if methods could be devised that do
not require such time dependent information.

Motivated by these considerations we propose a method with the novelty
that it uses only direct CP-violation and does not involve any use of time
dependent measurements. Potentially all types of B-mesons ($B_u$, $B_d$,
$B_s$) can be used at all B-facilities.  The method also allows for
experimental tests of the key theoretical approximations that are
involved. Specifically, we propose determination of the angles $\alpha$
and $\gamma$ through a study of the interference of tree and penguin
amplitudes in decays of B-mesons to two vector particles.

There have been several methods proposed to extract these angles.  For
$\alpha$ one can consider oscillation effects in $B^0\to\pi^+\pi^-$
although one must account for the penguin through isospin
analysis~\cite{alpha1} by observing $B^0\to\pi^0\pi^0$.  Since the
branching ratio to $\pi^0\pi^0$ is expected to be small and hard to
observe, it may be preferable to consider $B^0\to\pi^+\pi^-\pi^0$ where
one can also take advantage of resonance effects in the Dalitz
plot~\cite{alpha2};  however there may be problems in precise modeling of
the resonance structure. Another method for extracting $\alpha$ from the
interference of $u$-penguins with $t$-penguins in $B^0\to K^{(*)}K^{(*)}$
may overcome the disadvantages of the $2\pi$ and $3\pi$ final
states~\cite{londonKK} although the analogous $B_s$ decays are required
for the analysis. All three of these methods use time dependent CP
violation measurements of $B^0(\xba B^0)$ decays driven by the mixing in
that system while our method relies on direct (time independent) CP
violation only.

The angle $\gamma$ may be extracted at the $B$-factories through the
interference of $b\to u\xba c s$ and $b\to c\xba u s$~\cite{adsglw}.  In
the method we discuss here, we will obtain $\gamma$ through the
interference of the $b\to s$ penguin and the $b\to s u\xba u$ tree and
$\alpha$ from the interference of the $b\to d$ penguin and the $b\to d
u\xba u$ tree.

Our method requires measurements made in two separate decay modes.  One
mode of the form $B\to V_1V_2$, together with $\xba B \to \xba V_1 \xba
V_2$, ($B=B_u$, $B_d$ or $B_s$) which receives both tree and penguin
contributions and another mode $B\to V_3V_4$ which is a pure penguin.  
The procedure uses helicity amplitudes of $B\to V_1V_2$, which can be
inferred from decay distributions of the
vectors~\cite{german,palmer,chi_wolf,ddf_moments,as_krho}. By analyzing
this data together with the helicity amplitudes for $B\to V_3V_4$ one can
extract the tree-penguin weak phase difference.  From a $b\to d$ penguin
transition contributing to $B\to V_1V_2$, we deduce 
$|{V_{td}\over V_{ts}}
\sin\alpha|$
from which $|\sin\alpha|$ may be obtained once 
$|{V_{td}/V_{ts}}|$ is
known~\cite{vtd_note}; from a $b\to s$ penguin we can obtain
$|\sin\gamma|$. A list of such modes is shown in Table~\ref{table_one}.
Thus to implement this method to obtain $\alpha$ one needs to study one
mode from the $\alpha$-mode column and one mode from the ``pure penguin''
column. Likewise to obtain $\gamma$ one $\gamma$-mode and one pure penguin
mode is required. Using more than one mode from each column will, of
course, increase the analyzing power of the method.

Let us first discuss some of the tree penguin interference modes that are
suitable in our method for extracting $\alpha$.  In charged $B$ decays,
the relevant mode is $B^\pm\to\rho^\pm\omega$. In this case the tree graph
is color allowed which we find is advantageous in terms of statistical
power. Since the charge of the $\rho$ indicates the flavor of the initial
state, no other tagging is required. However, this mode has two $\pi^0$,
in the final state originating from decays of $\rho^\pm$ and $\omega$
perhaps making it difficult for hadronic machines.

From Table~\ref{table_one} we see see that for an analogous extraction of
$\alpha$ via decays of $B^0$ we may use any one of the modes
$B^0\to\rho^0\rho^0$, $\rho^+\rho^-$, $\omega\rho^0$ or $\omega\omega$.  
These modes require tagging at production and C-odd interference terms are
degraded somewhat by time integration.  Also the tree is color suppressed;
however, in the case of $\rho^0\rho^0$ it leads to a final state that does
not contain any $\pi^0$ and so may be easier to implement at hadronic $B$
machines. The mode $B^0\to\rho^+\rho^-$ has the advantage that any
electro-weak penguin (EWP) contribution is color suppressed.

Decays of $B_s$ usable for $\alpha$ are $B_s\to K^{*0}\rho^0$ or $B_s\to
K^{*+}\rho^-$. Both are self tagging, though in the first case only if
$K^{*0}\to K^+\pi^-$. The first mode also has the advantage that the final
state contains no $\pi^0$. The final state $K^{*+}\rho^-$, while suffering
from at least one $\pi^0$ in the final state, has the advantage that
possible contamination from the EWP is color suppressed.

Likewise, our method can also be used to extract $\gamma$ through the use
of $b\to s$ penguins.  In $B^0$ the candidate modes are $B^0\to K^{*+}
\rho^-$, $B^0\to K^{*0} \rho^0$ and $B^0\to K^{*0} \omega$; note that
$B^0\to K^{*0} \rho^0$ has the advantage of no $\pi^0$ in the final state.
For charged $B$'s again there are two such modes: $B^+\to K^{*+}\rho^0$
and $B^+\to K^{*+}\omega$, either one of which could be used to obtain
$\gamma$;  $B^+\to K^{*+}\rho^0$ is free of $\pi^0$.

Our method also needs input from one pure penguin mode; this should be a
$b\to s$ penguin since we need the component with an intermediate $u$
quark to be small.  For pure penguin modes we can use:  $B^+\to \phi
K^{*+}$, $B^0\to \phi K^{*0}$, $B^+\to K^{*0}\rho^+$, $B_s\to \phi \phi$
or $B_s\to K^{*0}K^{*0}$.  Likewise the decays $B^0\to\rho^0K^{*0}$ and
$B^0\to\omega K^{*0}$ can also be used since the tree contribution is
color and CKM suppressed. In Table~\ref{table_one} as well as listing
all the modes mentioned, we indicate which have color suppressed
EWP (underlined); which have color allowed tree
contributions (parentheses) and which have $\pi^0$-free final states
[square brackets].

\begin{table}[h]
\begin{center}
\caption{
The relevant $B\to VV$ modes originating from each kind of $B$ meson are
shown.  In the ``$\alpha$-mode'' column are tree-penguin interference
modes which are sensitive to $\alpha$; likewise the ``$\gamma$-modes'' are
sensitive to $\gamma$.  The ``pure penguin'' modes proceed through $b\to
s$ penguin processes only. The underlined modes have a color suppressed
EWP contribution; the modes enclosed in parentheses have
color allowed tree contributions while the modes enclosed in square
brackets have $\pi^0$ free final states.
\label{table_one}
}
\bigskip
\begin{tabular}{|c|c|c|c|}
\hline
$B$-meson		&
$\alpha$-mode		&
$\gamma$-mode		&
Pure Penguin		\\
\hline
\hline
$B^+$			
&
($\rho^+\omega$)	
&
[({$K^{*+}\rho^0$})],
($K^{*+}\omega$)
&
\tstrut
[$\phi K^{*+}$],			
$\underline{K^{*0}\rho^+}$
\\
\hline
$B^0$			&
(\underline{$\rho^+\rho^-$}),
[$\rho^0\rho^0$], 
&
(\underline{$K^{*+}\rho^-$})
&
[$K^{*0}\phi$]			
\\
&
$\omega\omega$, 
$\rho^0\omega$
&
&
\\
\hline
$B_s$			&
[$K^{*0}\rho^0$],
(\underline{$K^{*+}\rho^-$}),		&
see note~\cite{no_bs_gamma_modes}	&
[$\phi\phi$],
[\underline{$K^{*0}K^{*0}$}]
\\
&
$K^{*0}\omega$
&&
\\
\hline
\end{tabular}
\end{center}
\end{table}

In our analysis we will use the approximations that:  (1)~SU(3) is a valid
symmetry for penguin processes, (2)~the effects of the EWP are small,
(3)~the $q\bar q$ pair which arises in a strong penguin does not form a
single vector meson of the final state.  Recall that the EWP are assumed
to be small for other proposed methods~\cite{alpha1,alpha2,londonKK} as
well for extracting $\alpha$. Note also that for some of the modes in
Table~\ref{table_one} the EWP is color suppressed. Later we will discuss
ways to test each of these three approximations.

To illustrate our method, we now focus on $B^\pm\to\rho^\pm\omega$ (i.e.
$V_1V_2=\rho^+\omega$) where the discussion easily generalizes to the
other modes mentioned above.
The SM amplitude ${\fa}$ for this process can be written:

\begin{eqnarray}
\fa = T v_u + P_u v_u + P_c v_c + P_t v_t
\equiv \hat T v_u + \hat P v_t
\end{eqnarray}

\noindent where $T$ is the tree contribution to the amplitude, $P_i$ is
the penguin contribution due to the diagram with an internal quark of type
$u_i$ and $v_i=V^*_{ib}V_{id}$.  Unitarity of the CKM matrix allows us to
express the amplitudes in terms of $\hat T=T+P_u-P_c$ which we will refer
to as the corrected tree and $\hat P=P_t-P_c$, the corrected penguin.

For this mode it is useful to follow a (non-standard)  convention for the
weak phase where the corrected tree is zero. The amplitudes for
$B^+\to\rho^+\omega$ may then be written as:

\begin{eqnarray}
{\bf A}=
{\bf a} 
+ 
{\bf b} 
e^{i\alpha}
~~~{\rm where}~~~
{\bf a}= \hat T |v_u| 
~~~{\rm and}~~~
{\bf b}= \hat P |v_t| 
\end{eqnarray}

\noindent
and each of the bold face terms represent a set of helicity amplitudes.

\noindent The amplitude for the charge conjugate process in this
convention is ${\bf \bar A}= {\bf a} + \Pi {\bf b} e^{-i\alpha}$ where
$\Pi$ indicates parity.

Even knowing the full amplitudes in these decays we cannot hope to
determine $\alpha$.  This is evident writing the amplitude:

\begin{eqnarray}
{\bf A}={\bf c} + i {\bf b}\sin\alpha,
~~~
{\bf \xba A}={\bf c} - i \Pi {\bf b}\sin\alpha
~~~\Rightarrow~~~
{\bf b}\sin\alpha=({\bf A}-\Pi{\bf \bar A})/(2i)
\label{d_def}
\end{eqnarray}

\noindent where ${\bf c}={\bf a}+{\bf b}\cos\alpha$.  While we may extract
${\bf c}$ and ${\bf b}\sin\alpha$, we clearly can not separately obtain
$\sin\alpha$ without more information~\cite{isonote}.  To solve for
$\alpha$ then, we need two more pieces of information: (1) A means to fix
the relative weak phase between ${\bf A}$ and ${\bf \xba A}$. (2) A
separate normalization of the penguin contribution ${\bf b}$.

The measurement of pure penguin amplitudes will provide these remaining
two inputs.  The normalization of such a mode will clearly allow the
determination of $\sin^2\alpha$ while the ratios and phase differences
between the components of ${\bf b}$ can be matched with those determined
from eqn.~(\ref{d_def})  providing the condition needed to fix the
relative phase of 
${\bf A}$ and $\overline {\bf A}$.

Let us now consider how the necessary information about the components of
the amplitude may be extracted from the experimental data. For both the
$\omega$ and $\rho$ the polarization vector can be completely determined
in their rest frame by their decays. For $\rho^\pm\to\pi^\pm\pi^0$ we can
define the polarization vector as $\vec E_\rho=\vec P_{\pi^0}/|\vec
P_{\pi^0}|$ while for the $\omega$ we define $\vec E_\omega=\vec
P_{\pi^+}\times \vec P_{\pi^-}/|\vec P_{\pi^+}\times \vec P_{\pi^-}|$. Let
us introduce $\theta_1$ to be the polar angle between $\vec E_\rho$ and
the boost axis in the $\rho$ frame and likewise $\theta_2$ the polar angle
between $\vec E_\omega$ and its boost axis. In addition we introduce the
azimuthal angle $\Dd$ between $\vec E_\rho$ and $\vec E_\omega$ with
respect to the $\rho$ boost axis. It is convenient to describe the final
state in a helicity bases which contains the following states:  
$|R\ket=|+1\ket_\rho|+1\ket_\omega$,~ 
$|L\ket=|-1\ket_\rho|-1\ket_\omega$
and
$|0\ket=|0\ket_\rho|0\ket_\omega$. 
In the following we will rewrite the
transverse modes as parity eigenstates:  
$|S\ket=(|R\ket+|L\ket)/\sqrt{2}$ and $|P\ket=(|R\ket-|L\ket)/\sqrt{2}$
where the action of parity is $\Pi:|0\ket$,
$|S\ket\to+|0\ket$, $|S\ket$ and
$\Pi:|P\ket\to-|P\ket$.

The amplitudes for production of the $|0\ket$, $|S\ket$ and $|P\ket$
states 
respectively
may be expressed in the three component notation: ${\bf A}= \left
[A_0,~ A_S,~ A_P \right]$.  The angular distribution (see e.g.
\cite{chi_wolf})  for the two vector decay is $ d\Gamma/d\Phi
=\sum_{i=1}^{6} X_i f_i(\Phi) $ where $\Phi$ represents the phase space,
$d\Phi=dz_1dz_2d\Dd/(8\pi)$ and the basic distributions are given by:

\begin{eqnarray}
&&f_1=9z_1^2z_2^2;			
~~~~
f_2=(9/2)u_1^2u_2^2\cos^2\Dd;	
~~~~
f_3=(9/2)u_1^2u_2^2\sin^2\Dd;	
\nonumber\\
&&f_4=9\sqrt{2}z_1z_2u_1u_2\cos\Dd;
~~~
f_5=9\sqrt{2}z_1z_2u_1u_2\sin\Dd;	\nonumber\\
&&f_6= 9 u_1^2u_2^2\sin\Dd\cos\Dd 
\label{dist2}
\end{eqnarray}

\noindent with $z_i=\cos\theta_i$ and $u_i=\sin\theta_i$ ($i=1,2$). In
terms of the helicity amplitudes, the coefficients are thus:

\begin{eqnarray}
\begin{array}{lll}
X_1=|A_0|^2;&
X_2=|A_S|^2;&
X_3=|A_P|^2;
\\
X_4=Re(A_0A_S^*);& 
X_5=Im(A_0A_P^*);&
X_6=Im(A_sA_P^*);
\end{array}
\label{pars}
\end{eqnarray}

\noindent we have normalized the coefficients so that $Br=X_1+X_2+X_3$.

A convenient method to determine $X_i$ from the angular distribution is to
introduce a set of operators~\cite{ddf_moments} $g_i$ such that
$<g_i>=X_i/(X_1+X_2+X_3)$.  We will use the following set of 
operators with this property:

\begin{eqnarray}
&&g_1=(40f_1-5f_2-5f_3)/126,
~~
g_2=(265f_2-20f_1-85f_3)/504,
\nonumber\\
&&g_3=(265f_3-20f_1-85f_2)/504,
~~
g_{4,5}=25f_{4,5}/36,
~~
g_6=25f_6/72.
\label{moments}
\end{eqnarray}

Once an experimental determination of the observables $\{X_i\}$ has been
made, it is possible to determine the amplitudes up to the following
ambiguities: (1)~An overall phase, (2)~the transformation ${\bf A}\to
\Pi{\bf A}^*$.  Thus we have six observables determining three 
complex (amplitudes), or
equivalently six real parameters. The overall phase ambiguity means that
the remaining five amplitude parameters are over determined by the six
observables.  An unobservable phase may be applied to both ${\bf A}$ and
$\xba {\bf A}$ leaving one only the relative phase between the $B^+$ and
$B^-$ decays undetermined.  To be specific, we will take this weak phase
to be $\pp=\arg(A_0(B^-)A_0^*(B^+))$.  Once $\pp$ is determined, the
experimental observables in $B^+$ and $B^-$ decay will allow us to obtain
${\bf b}\sin\alpha$ from eqn.~(\ref{d_def}). To obtain $\sin\alpha$ we
therefore need a means to fix $\pp$ as well as information about ${\bf
b}$.

%
%

%
%

Consider now what may be learned from $B^0\to\phi K^*$. In the SM we
expect negligible CP violation in this mode (e.g. in~\cite{palmer} the CP
violation is estimated to be $O(1\%)$); thus $X_{1-4}(\phi
K^{*0})=X_{1-4}(\phi \xba K^{*0})$ and $X_{5,6}(\phi K^{*0})=-X_{5,6}(\phi
\xba K^{*0})$.  Within the SM and using our assumption that the gluon does
not fragment to a single vector meson, the amplitude ${\bf A}(\phi
K^{*0})=\sqrt{2}{\bf b}(V_{ts}/V_{td})$ hence the observables 
will be related to ${\bf b}$ by:

\begin{eqnarray}
&&{|b_0|^2}
=
q
{X_1({\cal F})},
~~
{|b_S|^2}
=
q
{X_2({\cal F})},
~~
{|b_P|^2}
=
q
{X_3({\cal F})},
~~
{Re(b_0b_S^*)}
=
q
{X_4({\cal F})},
\nonumber\\
&&
{Im(b_0b_P^*)}
=
q
{X_5({\cal F})},
~~
{Im(b_Sb_P^*)}
=
q
{X_6({\cal F})},
\label{final_eqn}
\end{eqnarray}

\noindent where ${\cal F}=\phi K^*$ and $q=|V_{td}/V_{ts}|^2/2$. We are
now in a position to determine $\pp$ and $\sin^2\alpha$ since
eqn.~(\ref{final_eqn}) gives six equations in only two unknowns (i.e.
$\pp$ and $\alpha$) when combined with eqn.~(\ref{d_def}).

We now estimate the precision with which one may extract $\sin^2\alpha$
using this method.  The statistical errors will, of course, depend on the
number of $B$ mesons which are available. To quantify this for a given
decay mode $B\to X$, let us define $\hat
N_B=(number~of~B)(acceptance~for~X)$ for each type of $B$-meson. For
concreteness, we consider $\hat N_B=10^8$ and $5\times 10^8$ in our
calculations.

The pure penguin modes $B^+\to\phi K^{*+}$ and $B^0\to\phi K^{*0}$ have in
fact recently been observed at BaBar~\cite{cleo_Kphi,babarpeng} with
$Br(B^+\to \phi K^{*+})$ = $(9.7^{+4.2}_{-3.4}\pm1.7)\times10^{-6}$ and
$Br(B^0\to \phi K^{*0})$ = $(8.6^{+2.8}_{-2.4}\pm1.1)\times10^{-6}$.  In
our sample calculation we will take this branching ratio to be $10^{-5}$.  
Estimating the corrected $b\to d$ penguin contributions, we get
$Br_{penguin}(B^-\to\rho^-\omega)$ $\approx$ $2|V_{td}/V_{ts}|^2 Br(B\to
\phi K^*)\approx 8\times 10^{-7}$ where~\cite{CKMparadgim}
$|V_{td}/V_{ts}|\approx 0.2$.  The tree graph for this final state has a
similar topology to the observed~\cite{cleo_Kpi,babar_0105061}
$B^+\to\pi^+\pi^0$ decay modes with branching ratios of $\sim 5\pm 2
\times 10^{-6}$.  The branching ratio to a two vector final state should
be somewhat larger because of the additional helicity states. Here we will
assume that this will increase the branching ratio by roughly a factor of
3 as is the case for $Br(B^+\to\rho^+ D^{*0})/Br(B^+\to\pi^+D^{0})$ so
that $Br_{tree}(B^+\to\rho^+\omega)\approx 15\times 10^{-6}$~\cite{pdb}.

The precision for extracting $\sin^2\alpha$ will depend on the
distribution of the amplitudes between the various helicity states. As an
example, which we will call Case 1, we will take the toy model for the
amplitudes given by:

\begin{eqnarray}
{\bf a}\propto
\left [1,1,-1\right ]
~~~~
{\bf b}\propto
\left [ 0.23,0.23,0.23\right ]
\label{toy_modelA}
\end{eqnarray}

\noindent which has the correct ratio of total rates.  In this toy model
the amplitudes have no relative phase from final state interactions.  
Case 2 will be the corresponding results averaged over helicity
distributions and phases subject only to the condition that $|{\bf a}|$
and $|{\bf b}|$ are fixed.

To simulate the reconstruction of $\sin^2\alpha$ from experimental data we
will calculate the observables $X_i$ for each mode and estimate the
correlated errors assuming that these are measured using the operators
$g_i$.  The tree-penguin mode give us $A$ and $\xba A$ up to a relative
phase. We can thus use eqn.~(\ref{d_def}) to get ${\bf b}\sin\alpha$ so
that eqn.~(\ref{final_eqn})  further constrains the fit and a minimum
value of $\chi^2$ can be obtained for each value of $\sin^2\alpha$.  The
dependence of $\chi^2$ on $\sin^2\alpha$ thus allows us to obtain the
statistical error $\Delta \sin^2\alpha$ given in Table~\ref{table_two}
where we take $\alpha=30^\circ$ and $60^\circ$. In the last line of this
Table, we consider a similar exercise for a mode sensitive to $\gamma$.

Table~\ref{table_two} gives results for the tree penguin modes: $B^+\to
\rho^+\omega$ where the tree is color allowed, $B^0\to \rho^0\omega$ where
the tree is color suppressed and a combined fit with both modes taken
together. We also consider the $B_s$ mode, $B_s\to K^{*+}\rho^-$. In each
case we take $\hat N_B=10^8$ and $5\times 10^8$ and give results for Cases
1 and 2.

\begin{table}[h]
\begin{center}
\caption{
The error in $\sin^2\alpha$ or $\sin^2\gamma$ assuming the central value
of $\alpha$, $\gamma$ = $30^\circ$ or $60^\circ$. The results under
Case 1 use the helicity distribution given in eqn.~(\ref{toy_modelA})
while Case 2 is a 
Monte Carlo
average both over helicity distribution and 
over phases keeping
the branching ratios fixed. In each entry the first number is for $\hat
N_B=10^8$ while the second is for $\hat N_B=5\times 10^8$.
\label{table_two}
}
\bigskip
\begin{tabular}{||c|c||c|c||c|c||}
\hline
& & 
\multicolumn{2}{c||}{Case 1} &
\multicolumn{2}{c||}{Case 2}
\\
Tree-Penguin		&
CKM quantity		&
$30^\circ$		&
$60^\circ$     	        &
$30^\circ$		&
$60^\circ$     	        
\\
\hline
\hline
$B^+\to\rho^+\omega$    	&
$\sin^2\alpha$          	&
.065; .029 			&
.112; .050			&
.063;.033 			&
.115;.062			\\ 			
\hline
$B^0\to\rho^0\omega$    	&
$\sin^2\alpha$          	&
.220;	.118			&
.221;	.107			&
.095; .051			&
.114;.060 			\\ 			
\hline
$B^+\to\rho^+\omega$ \& $B^0\to\rho^0\omega$   &
$\sin^2\alpha$          	&
.056; .025			&
.101; .045			&
.055;.023 			&
.081;.040			\\ 			
\hline
$B_s\to K^{*+}\rho^-$    	&
$\sin^2\alpha$          	&
.088; .039			&
.151; .068			&
.098;.039 			&
.146;.072			\\ 			
\hline
$B^+\to K^{*+}\rho^-$    	&
$\sin^2\gamma$          	&
.054; .032			&
.069; .049			&
.099; .054 			&
.095; .057			\\ 			
\hline
\end{tabular}
\end{center}
\end{table}

Let us now consider how the validity of the approximations used in this
method may be experimentally verified.  The validity of SU(3) for penguins
may be tested by comparing the various pure penguin modes considered.  If
this symmetry is exact, the magnitude and relative phases of all the $b\to
s$ pure penguin modes should be the same.

%
%

The assumption that EWPs do not play a large role is required in most
isospin analysis~\cite{alpha1, alpha2, londonKK} since the EWPs have the
isospin properties of the tree but the weak phase of the penguin so it
cannot be cleanly separated from the tree without additional information.
Its presence may be specifically checked in the decay
$B^\pm\to\rho^\pm\rho^0$. The final state must have total isospin $I=2$ so
in the SM only the tree and EWP can contribute.  A significant EWP
contribution which has a relative weak phase of $\alpha$ with respect to
the tree would thus lead to CP violation. There are six possible CP
violating observables that can be obtained from the distributions:  
$X_{1-4}(\rho^+\rho^0)-X_{1-4}(\rho^-\rho^0)$ and
$X_{5,6}(\rho^+\rho^0)+X_{5,6}(\rho^-\rho^0)$.  Note that the first four
are even under naive time reversal~\cite{physrep}, $T_N$, thus require a
rescattering phase while the last two are $T_N$ odd and do not. The
analogous test for the EWP can also be performed in the decay
$B^\pm\to\pi^\pm\pi^0$ where an EWP could generate a partial rate
asymmetry.  The possible advantage of using the $\rho^\pm\rho^0$ final
state is that there are six CP odd signals rather than one in
$\pi^\pm\pi^0$.

In the $K\pi$, $K^*\pi$, $K\rho$ and $K^*\rho$ systems one can also check
for EWP effects by looking for CP violation with isospin 
properties
inconsistent
with tree penguin interference~\cite{as_krho,as_kpi}. 
In the $K\pi$ system the absence of EWP implies:

\begin{eqnarray}
2  \Delta(B^-\to K^-\pi^0)
-  \Delta(B^-\to \xba K^0\pi^-)
-  \Delta(\xba B\to \xba K^-\pi^+)
+2 \Delta(\xba B\to \xba K^0\rho^0)=0
\label{sumrule}
\end{eqnarray}

\noindent
where $\Delta(X)$ is the partial rate asymmetry for the reaction $X$. 
Eqn.~(\ref{sumrule}) also applies to 
$K\rho$ and $K^*\pi$ while for 
$K^*\rho$ it applies separately to $X_1$, $X_2$ and
$X_3$.

It is also possible to reduce the potential impact of the EWP by choosing
modes in which their contribution will be color suppressed. In particular,
the tree penguin interference processes $B_s\to K^{*+}\rho^-$ and $B_0\to
\rho^+\rho^-$ will color suppress the EWP as will the
pure penguin process $B_s\to K^{*0} K^{*0}$.

The assumption that the gluon does not fragment into a single vector meson
is true at lowest order in perturbation theory by color conservation.  We
can estimate how well it holds when QCD corrections are introduced,
through renormalization group improved perturbation theory. If we define
${\bf b^\prime}$ to be the penguin contribution where the quarks from the
gluon form a single meson, the ratios of the contributions should be:

\begin{eqnarray}
|{\bf b^\prime}|/|{\bf b }|
\approx
{
(3C_3+C_4+3C_5+C_6)
/
(C_3+C_4+C_5+C_6)
}
\end{eqnarray}

\noindent where we have used the notation of~\cite{buras_rg}.  With the
numerical results from that paper we find that this ratio is about 0.04
for various values of $\Lambda_{QCD}$ both in naive dimensional
regularization and in the ~'t-Hooft---Veltman schemes at next to leading
order.


To summarize, we have shown that unitarity angles $\alpha$ and $\gamma$
can be extracted by a method that is rather unique in that it only
uses direct CP violation and does not need any time dependent
measurements. We gave several examples of modes which can be used and also
discussed how experimentally the underlying theoretical approximations can
be tested. Since these angles are fundamental parameters of the 
SM, and their precise determinations could lead to its crucial tests,
the importance of measuring them in several different ways can hardly be
over emphasized.

%

\bigskip

This research was supported in part by US DOE Contract Nos.\
DE-FG02-94ER40817 (ISU); DE-AC02-98CH10886 (BNL).



\begin{thebibliography}{99}
\bibitem{vtd_note} Using the fits in~\cite{CKMparadgim}
$|V_{td}/V_{ts}|=0.183\pm 0.015$. Future measurement of the rate of $B_s$
oscillation at the Tevatron or LHC should determine this ratio with much
greater accuracy.
\bibitem{no_bs_gamma_modes} In the case of $B_s$, there are no analogous
self tagging modes which may be used;  for instance, the decay mode
$B_s\to K^{*+}K^{*-}$ is not useful since it cannot be flavor tagged.
\bibitem{isonote} Note that because the tree always has a component with
the same isospin properties as the penguin, the consideration of isospin
related processes alone will therefore not help to separate ${\bf
b}\cos\alpha$ from ${\bf a}$.
\end{thebibliography}
\end{document}